\documentclass[double]{article}
\usepackage[bookmarks = true, pdfpagemode = None, pdfstartview = FitH,colorlinks = true, citecolor = black, linkcolor = black, urlcolor = black]{hyperref} 
\input{Qcircuit.tex}
\usepackage{cases}

\usepackage{amssymb}
\newtheorem{theorem1}{\textbf{Theorem}}

\newtheorem{definition}{\textbf{Definition}}

\newtheorem{conjecture}{\textbf{Conjecture}}

\newtheorem{Requirement}{\textbf{Quantum Test Requirement}}

\newtheorem{proof}{\textbf{Proof}}

\title{Fault testing quantum switching circuits}
\author{Jacob D Biamonte and Marek Perkowski\footnote{JDB and MP are with Portland State University,
Portland, Oregon 97201, USA.  JDB present address: Oxford University Computing Laboratory, Wolfson Building, Parks Road, Oxford, OX1 3QD, UK.}}
\begin{document}
\maketitle

\begin{abstract}
Test pattern generation is an electronic design automation tool that
attempts to find an input (or test) sequence that, when applied to a
digital circuit, enables one to distinguish between the correct
circuit behavior and the faulty behavior caused by
particular faults. The effectiveness of this classical method is
measured by the fault coverage achieved for the fault model and the
number of generated vectors, which should be directly proportional
to test application time. This work address the quantum process
validation problem by considering the quantum mechanical adaptation
of test pattern generation methods used to test classical circuits.
We found that quantum mechanics allows one to execute multiple test
vectors concurrently, making each gate realized in the process act
on a complete set of characteristic states in space/time complexity
that breaks classical testability lower bounds.
\end{abstract}

\section{Introduction}\label{sec:intro}
Classically, both
the fault models considered and the test inputs are localized (not
entangled). Form this fact and other reasons, it is not immediately
apparent how methods developed in classical test theory could be put
to use when testing quantum mechanical switching networks.  Here ones views a
quantum circuit as being a description of the actions on, and the
interactions between qubits. To avoid brute force testing, a method to
test these time dependent connections as well as the logical
operation of each gate realized in a process is presented. This is
done by considering a logical set of failure models designed to
drive a switching network to it's bounds of operation and assure
main aspects of individual gate functionality.

It is theoretically interesting to combine classical test theory
with quantum effects such as entanglement. For instance, quantum states can
be designed to execute multiple test vectors concurrently. In
addition quantum systems are reversible and any
reversible system preserves information. This means that a
reversible system preserves the probability that additional
information may be present~\cite{Zurek:RSIPS:84}.  The additional
information present could be used to detect the presence of a fault. In a faulty reversible circuit, the
probability of detection for quantum fault $f$ observable with
$\hat{A}$ is solely related to the probability of
$f$'s presence (see $\S$~\ref{sec:Fault Models}). For instance, one
may develop a test set, consisting of input vectors and
corresponding observables, that separates a circuit from all
considered faults. Based on the information conservative properties
of reversible systems~\cite{Zurek:RSIPS:84}, this test set can be
applied multiple times to detect the probabilistic version of the
considered faults.

We consider extending methods developed to test
classical circuits, and so present tests designed
to separate an oracle from one containing a given set of considered
faults~\cite{Biamonte:05}. Classically, the testability of the
circuit class comprising the oracle has already received much
attention after the $1972$ paper by Sudhakar M. Reddy~\cite{reddy}.
This paper presents a quantum mechanical
generalization of this and several other classical
methods~\cite{kautz:tff:67,reddy,McCluskey:SFvsD:00}.

\textbf{Structure of the paper:} Sec.~\ref{subsec:Constructing Binary
Quantum Networks} gives an introduction to
oracle construction. Sec.~\ref{sec:Fault Models} discusses the
quantum fault models used in this study. The Quantum Test Algorithm
is presented in Sec.~\ref{sec:Principle Method} followed by the
conclusion in Sec.~\ref{sec:Conclusion}.

\subsection{Constructing Quantum Oracle Search Spaces}\label{subsec:Constructing Binary Quantum
Networks}

 Any Boolean equation may be
uniquely expanded to the Positive Polarity Reed-Muller Form
(PPRM)~\cite{reddy} as:\setlength{\arraycolsep}{0.140em}
\begin{eqnarray}\label{eqn:rm_exp}
&&f(x_1,x_2,...,x_k) = c_0\oplus c_1 x_1^{\sigma_1}\oplus c_2 x_2^{\sigma_2}\oplus \cdots \oplus c_n x_n^{\sigma_n}\oplus\nonumber\\
&&~~~~~~~~c_{n+1}x_1^{\sigma_1} x_n^{\sigma_{n}}\oplus \cdots \oplus
c_{2k-1}x_1^{\sigma_1} x_2^{\sigma_2},...,x_k^{\sigma_k},
\end{eqnarray}where selection variable $\sigma_i\in \{0,1\}$, literal $x_i^{\sigma_i}$ represents a variable or its negation and any $c$ term labeled
$c_0$ through $c_j$ is a binary constant $0$ or $1$. 

\emph{Example:}
\begin{eqnarray}\label{eqn:example PPRM}
f(x_1,x_2,x_3,x_4)&=&1\oplus x_1\oplus x_2\oplus x_3 \oplus x_3 x_4\oplus x_1 x_3 x_4\oplus \nonumber\\
&&~x_1 x_2 x_3\oplus x_2 x_3 x_4
\end{eqnarray}

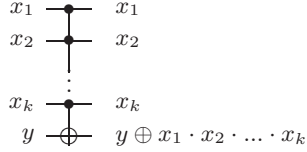
\begin{figure}[t]
\small{\centerline{  \Qcircuit @C=.5em @R=.5em @!R {
\lstick{x_1} & \ctrl{1} & \qw & \rstick{x_1}\\
\lstick{x_2} & \ctrl{1} & \qw & \rstick{x_2}\\
        & \mbox{\vdots} & &&&&&&&&\\
\lstick{x_k} & \ctrl{1} & \qw & \rstick{x_k}\\
\lstick{y} & \targ & \qw & \rstick{y\oplus x_1\cdot x_2\cdot
...\cdot x_k}} }} \caption{$k-$CN Gate Realizing $y\oplus x_1\cdot
x_2\cdot ...\cdot x_k$ on the $(k+1)^{th}$
qubit.}\label{cir:Arbitrary Quantum controlled-NOT Gate}
\end{figure}

Each term in the expansion of Eqn.~\ref{eqn:example PPRM} is called
a product term~\cite{UgurHP}, and each variable $x_i$ a literal.  
For example, $x_3\cdot x_4$ is a product term,
with literals $x_3$ and $x_4$ (constant $1$ is not considered to be
a product term). Each product term for a given PPRM expansion is
realized by an arbitrary quantum controlled-NOT gate ($k-$CN) given
in Fig.~\ref{cir:Arbitrary Quantum controlled-NOT Gate}.  Repeating
this procedure for each product term in Eqn.~\ref{eqn:example PPRM}
and sequencing the gates leads to the network realization given in
Fig.~\ref{cir:Quantum Network Realization of Example
Function}. Above each gate is the label $p_i$, $p$ refers to a
product term in the expansion of Eqn.~\ref{eqn:example PPRM}, and
$i$ the index used to label the seven products. This example will be
used again so it is stated explicitly that $p_0$ corresponds to
$x_1$, $p_1$ to $x_2$, $p_3$ to $x_3$, $p_3$ to $x_3 x_4$, $p_4$ to
$x_1 x_3 x_4$, $p_5$ to $x_1 x_2 x_3$ and finally $p_6$ to $x_2 x_3
x_4$.



In many quantum algorithms, after Boolean function $f$ is constructed by means of a $k-$CN
network it is \textit{placed} in a black box oracle ($\mathcal{O}$). The
bottom $(k+1)^{th}$ bit contains the realization of $f$ to be read
at the box's right. The top $k$ inputs to the box begin in state
$\ket{0}$ and the $(k+1)^{th}$ input (target) qubit starts in state
$\ket{1}$. The Hadamard operation $H^{\otimes(k+1)}$ is applied
placing the input query in a superposition of all $2^{(k+1)}$
classical states. Generally the black box takes as input:
\begin{equation}\label{ean:h}
H^{\otimes (k+1)}:\ket{0}^{\otimes k}\otimes\ket{1}\longrightarrow
(\ket{0}+\ket{1})^{\otimes k}\otimes(\ket{0}-\ket{1})
\end{equation}
Inside the black box all of the targets act on state $\ket{-}$ (an
eigenvector of the $k-$CN gate) and the top $k$ qubits remain in a
superposition.  The true \emph{minterms} are inputs that make a
Boolean function evaluate to $1$ where false \emph{minterms}
evaluate to $0$. Each term in the superposition on the top $k$ bits
representing a true minterm in the switching function $f$ realized
in the oracle will be appended with a negative (relative) phase. The
phase of states that do not represent true minterms are left
invariant. This is seen by examining the truth table from
Fig.~\ref{fig:truth_table}. The action of an oracle
$\mathcal{O}$,\footnote{$\mathcal{O}$ is sometimes represented as,
$\sum_x(-1)^{f(x)}\ket{x}\bra{x}$.} realizing a binary function
$f(x_1,x_2,...,x_k)$, is represented by the transform:
\begin{equation}
\mathcal{O}:\ket{k}\otimes\ket{-}\longrightarrow
(-1)^{f(k)}\ket{k}\otimes\ket{-}.
\end{equation}



\begin{figure}[t]
\small{\centerline{  \Qcircuit @C=.8em @R=.5em @!R {
             & \mbox{$p_0$} & \mbox{$p_1$} & \mbox{$p_2$} & \mbox{$p_3$} & \mbox{$p_4$} & \mbox{$p_5$} & \mbox{$p_6$} &&&&&&&&\\
\lstick{x_1} & \ctrl{4}     & \qw          & \qw          & \qw          & \ctrl{2}     & \ctrl{1}     & \qw          & \qw      & \rstick{x_1}\\
\lstick{x_2} & \qw          & \ctrl{3}     & \qw          & \qw          & \qw          & \ctrl{1}     & \ctrl{1}     & \qw      & \rstick{x_2}\\
\lstick{x_3} & \qw          & \qw          & \ctrl{2}     & \ctrl{2}     & \ctrl{2}     & \ctrl{2}     & \ctrl{1}     & \qw      & \rstick{x_3}\\
\lstick{x_4} & \qw          & \qw          & \qw          & \ctrl{1}     & \ctrl{1}     & \qw          & \ctrl{1}     & \qw      & \rstick{x_4}\\
\lstick{y=1} & \targ        & \targ        & \targ        & \targ &
\targ & \targ & \targ    & \qw      & \rstick{1\oplus
f(x_1,x_2,x_3,x_4)}} }} \caption{Quantum Network Realization of
Eqn.~\ref{eqn:example PPRM} built from arbitrary $k$-CN gates as
shown in Fig.~\ref{cir:Arbitrary Quantum controlled-NOT Gate}.  The
truth table of this oracle is given in Fig.~\ref{fig:truth_table}.
}\label{cir:Quantum Network Realization of Example Function}
\end{figure}
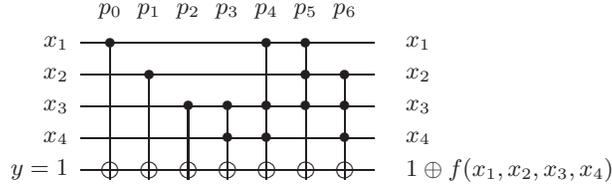
\begin{figure}[h]\centerline{
\small{
\begin{tabular}{lr}
  \begin{tabular}[h]{c||cccc||c}
\emph{phase~~state~~~~~}&$x_1$&$x_2$&$x_3$&$x_4$&$f$\\
\hline
$+~~~\ket{0000}$&0&0&0&0&0\\
$+~~~\ket{0001}$&0&0&0&1&0\\
$-~~~\ket{0010}$&0&0&1&0&1\\
$+~~~\ket{0011}$&0&0&1&1&0\\
$-~~~\ket{0100}$&0&1&0&0&1\\
$-~~~\ket{0101}$&0&1&0&1&1\\
$+~~~\ket{0110}$&0&1&1&0&0\\
$+~~~\ket{0111}$&0&1&1&1&0\\
$-~~~\ket{1000}$&1&0&0&0&1\\
$-~~~\ket{1001}$&1&0&0&1&1\\
$+~~~\ket{1010}$&1&0&1&0&0\\
$+~~~\ket{1011}$&1&0&1&1&0\\
$+~~~\ket{1100}$&1&1&0&0&0\\
$+~~~\ket{1101}$&1&1&0&1&0\\
$+~~~\ket{1110}$&1&1&1&0&0\\
$-~~~\ket{1111}$&1&1&1&1&1\\
  \end{tabular}
\end{tabular}}}
\caption{Oracle Truth Table for Eqn.~\ref{eqn:example PPRM}
implemented by the network in Fig.~\ref{cir:Quantum Network
Realization of Example Function}: Boolean function $f$ is
implemented quantum mechanically. Each of the $2^k$ terms in a
superposition input that evaluate to \emph{logic-one} will be marked
with a negative phase (also shown in Eqn.~\ref{eqn:t3-2}, in
Sec.~\ref{sec:Principle Method}).}\label{fig:truth_table}
\end{figure}

\section{Gate Level Quantum Fault Models}\label{sec:Fault Models}

\begin{figure}[h]\small{
\centerline{
    \Qcircuit @C=1em @R=.2em {
 &                  & \mbox{$_1$}    &                & \mbox{$_2$}    & \\\\\\
 & \lstick{\ket{i_0}} & \qswap         & \ctrl{6}       & \qswap         &  \measuretab{m_0}\\\\\\\\
 &                  & \mbox{$_3$}    &                & \mbox{$_{4}$} & \\\\
 & \lstick{\ket{i_1}} & \qswap         & \ctrl{4}       & \qswap         &  \measuretab{m_1}\\\\\\
 &                  & \mbox{$_{5}$} &                & \mbox{$_{6}$} &\\
 & \lstick{\ket{i_2}} & \qswap         & \targ          & \qswap         &  \measuretab{m_2}}}}
 \caption{\label{tofgate2}
 $2-$CN gate with error locations.}
 \end{figure}
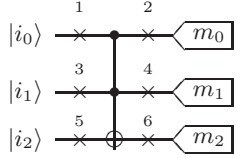

Consider the single stage circuit shown in
Fig.~\ref{tofgate2}. The numbered locations of possible gate
external faults are illustrated by placing an "$\times$" on the line
representing a qubits time traversal and here, the gate, initial
states ($\ket{i_0}, \ket{i_1}, \ket{i_2}$) and measurements ($m_0,
m_1, m_2$) may also contain errors.  

\begin{definition}\label{Error Location}
    Error/Fault Location: The wire locations between stages as well as any node, gate initial state or measurement in a given network $($see Fig.~\ref{tofgate2}$)$.
\end{definition}

\begin{definition}\label{def:Single Quantum Fault Model}
    Quantum Single Fault Model: For simplification the "quantum single fault model" is assumed in this work.
    In the single fault model, test plans are optimized for all considered faults assuming that only a single failure perturbs the quantum circuit
    exclusively.  Multiple faults will accumulate
    and be detected, but the single fault model makes it much
    easier to develop test plans.
\end{definition}

\begin{conjecture}\label{conjecture:qsfm}
A test set designed to detect all considered single errors will
detect and sample the accumulated impact of multiple errors at
multiple locations.
\end{conjecture}

The following definitions are used to define some of the fault types
considered in this work.  Complete fault coverage occurs after a
test set has determined that the considered fault(s) are not present in a given circuit.  

\begin{definition}\label{def:Pauli Single Fault Model}
    Pauli Fault Model: The addition of an
    unwanted Pauli matrix in a quantum network, at any error location and with placement probability $p$.
    The Pauli matrices are given in Eqn.~\ref{eqn:sigma_x},~\ref{eqn:sigma_y}
    and~\ref{eqn:sigma_z}.
\end{definition}

\begin{equation}\label{eqn:sigma_x}
\sigma_{x}=\ket{1}\bra{0}+\ket{0}\bra{1}
\end{equation}
\begin{equation}\label{eqn:sigma_y}
 \sigma_{y}=i\ket{0}\bra{1}-i\ket{1}\bra{0}
\end{equation}
\begin{equation}\label{eqn:sigma_z}
\sigma_{z}=\ket{0}\bra{0}-\ket{1}\bra{1}
\end{equation}

\begin{definition}\label{def:Single Rotation Initialization Fault}
    Initialization Error: A qubit that
    statistically favors correct preparation in one basis state over the other.
\end{definition}

\begin{definition}\label{def:Measurement Single Fault Model}
    Measurement Fault Model: A single functional
    measurement gate is replaced with a faulty measurement gate that statistically favors returning \emph{logic-zero} or a \emph{logic-one}.
\end{definition}

\begin{Requirement}\label{Requirement:bit flip}
    A bit flip ($\sigma_x$ or $\sigma_y$) at any error location must be detectable.~$\blacksquare$
\end{Requirement}

\begin{Requirement}\label{Requirement:phase flip}
    A phase flip ($\sigma_z$ or $\sigma_y$) at any error location must be detectable.~$\blacksquare$
\end{Requirement}

\begin{Requirement}\label{Requirement:initialization fault model}
    Each qubit must be initialized in both basis states $\ket{0}$ and $\ket{1}$.~$\blacksquare$
\end{Requirement}

\begin{Requirement}\label{Requirement:phase1}
    With the target acting on state $\ket{-}$: Each gate must be
    shown to attach a relative phase to arbitrary activating state $\ket{a}$ with both positive and negative eigenvalues.
    Furthermore, each gate must be shown not to attach a relative phase to arbitrary non-activating
    state $\ket{n}$ with both positive and negative eigenvalues.  The target state must remain
    globally invariant under both $\ket{a}$ and
    $\ket{n}$.~$\blacksquare$
\end{Requirement}

\begin{Requirement}\label{Requirement:lost phase}
    With the target acting on state $\ket{+}$: relative phase must be shown not to change under arbitrary activating
    state $\ket{a}$ with both positive and negative eigenvalues.
    Furthermore, relative phase must not change under arbitrary non-activating
    state $\ket{n}$ with both positive and negative eigenvalues.~$\blacksquare$
\end{Requirement}

\begin{Requirement}\label{Requirement:fadedcontrol}
    For the target acting separately on basis state $\ket{0}$ and $\ket{1}$:
    All controls in a gate must be activated concurrently.  Furthermore, each
    control must be addressed with a non-activating state.~$\blacksquare$
\end{Requirement}

\begin{Requirement}\label{Requirement:gates act on full basis}
    Each target must separately act on basis state inputs $\ket{0}$ and $\ket{1}$.~$\blacksquare$
\end{Requirement}

\begin{Requirement}\label{Requirement:Measurement}
    Each qubit must be measured in both \emph{logic-zero} and
    \emph{logic-one} states.~$\blacksquare$
\end{Requirement}

\subsection{Conclusions based on the Gate Level Fault
Models}\label{par:conGate}

In practice, the choice of the fault model
will be determined by a particular quantum circuit technology, as
well as how the circuit will be used.  In this work the functional
use of $k-$CN networks are oracle search spaces.  In this setting, 
any $k-$CN gate exhibits twelve, functionally
distinct actions.  

\begin{theorem1}\label{theoremclassical}
    A quantum $k-$CN gate is capable of four characteristic classical
    operations. (By characteristic it is meant that all other
    operations are variants of this basic set.)
\end{theorem1}

\begin{proof}
The gate is able to act on a $\ket{0}$ and a $\ket{1}$ state when
all controls are set. The two remaining functions are simply
to act on $\ket{0}$ and $\ket{1}$ when one or more control(s) is
addressed with a non-activating state. There are $2^k-1$ input states that do not
activate the gate, but these inputs all probe the \emph{off}
function. 

Similarly, each control has two logical functions. The
first is to be addressed with a logical $\ket{0}$ and the second is
to be addressed with a $\ket{1}$.~$\blacksquare$
\end{proof}

\begin{figure}[h]\[\renewcommand{\arraycolsep}{.04in}
\begin{array}{ccccccc}
v_0 \rightarrow & 0 & 0 & 0 & 0 & \cdots & 1 \\
v_1 \rightarrow & 0 & 0 & 0 & 0 & \cdots & 0 \\
v_2 \rightarrow & 1 & 1 & 1 & 1 & \cdots & 1 \\
v_3 \rightarrow & 1 & 1 & 1 & 1 & \cdots & 0 \\
\end{array}
\]\caption{Classical test vectors ($v_0$,
$v_1$, $v_2$, $v_3$) acting on binary basis vectors $\{0, 1\}$ with
the gate first off ($v_0, v_1$) and then on ($v_2, v_3$). The
rightmost bit in the figure is applied to the $(k+1)^{th}$
bit.}\label{fig:Classical test vectors v_1-v_4}\end{figure}

\begin{figure*}[t]\begin{center}
\small{\begin{tabular}{rl|rl}
  \emph{Minterm~~~} & \emph{~~Target State} & \emph{Minterm~~~} & \emph{~~Target State} \\
  \hline\hline
  $e^{+i\phi}\ket{true}$  & $\left(\ket{0}+e^{+i\varphi}\ket{1}\right)$ & $e^{+i(\phi+\varphi)}\ket{true}$ & $\left(\ket{0}+e^{+i\varphi}\ket{1}\right)$ \\
  $e^{-i\phi}\ket{true}$  & $\left(\ket{0}+e^{+i\varphi}\ket{1}\right)$ & $e^{-i(\phi-\varphi)}\ket{true}$ & $\left(\ket{0}+e^{+i\varphi}\ket{1}\right)$ \\
  $e^{+i\phi}\ket{false}$ & $\left(\ket{0}+e^{+i\varphi}\ket{1}\right)$ & $e^{+i(\phi)}\ket{false}$ & $\left(\ket{0}+e^{+i\varphi}\ket{1}\right)$ \\
  $e^{-i\phi}\ket{false}$ & $\left(\ket{0}+e^{+i\varphi}\ket{1}\right)$ & $e^{-i(\phi)}\ket{false}$ & $\left(\ket{0}+e^{+i\varphi}\ket{1}\right)$ \\
  \hline
  $e^{+i\phi}\ket{true}$  & $\left(\ket{0}+e^{-i\varphi}\ket{1}\right)$ & $e^{+i(\phi-\varphi)}\ket{true}$ & $\left(\ket{0}+e^{-i\varphi}\ket{1}\right)$ \\
  $e^{-i\phi}\ket{true}$  & $\left(\ket{0}+e^{-i\varphi}\ket{1}\right)$ & $e^{-i(\phi+\varphi)}\ket{true}$ & $\left(\ket{0}+e^{-i\varphi}\ket{1}\right)$ \\
  $e^{+i\phi}\ket{false}$ & $\left(\ket{0}+e^{-i\varphi}\ket{1}\right)$ & $e^{+i\phi}\ket{false}$ & $\left(\ket{0}+e^{-i\varphi}\ket{1}\right)$ \\
  $e^{-i\phi}\ket{false}$ & $\left(\ket{0}+e^{-i\varphi}\ket{1}\right)$ & $e^{-i\phi}\ket{false}$ & $\left(\ket{0}+e^{-i\varphi}\ket{1}\right)$ \\
  \hline\hline
\end{tabular}}
\caption{A $k-$CN Gate Truth Table (Case: 2 top, Case: 1 bottom):
Illustrating all of the different possible gate actions for
orthogonal setting of variables $\phi$ and $\varphi$. A $\ket{true}$
minterm activates the gate, any $\ket{false}$ minterm does not.}
\label{tab:CN variables Truth Table}
\end{center}\end{figure*}

Provided the state of the top $k$ bits is some equal superposition
and the target of the gate acts on a state with the following form:
$\ket{0}+e^{\pm i\varphi}\ket{1}$.  Under this condition, the inputs
to a $k-$CN gate are expressed as:
\begin{equation}\label{eqn:genstate}
\ket{\psi_{in}}\longrightarrow
\left[\sum_{x~=~0}^{2^k-1}w_x\ket{x}\right]\otimes(\ket{0}+e^{\pm
i\varphi}\ket{1}),
\end{equation}
where $w_x = e^{\pm i\phi}$.  Similarly, as in the case of
Theorem~\ref{theoremclassical}, certain operations define the gate's
function. 

The arbitrary quantum superposition state defined in
Eqn.~\ref{eqn:genstate} allows one to consider each input as a
separate state.  In the column denoted minterm from Fig.~\ref{tab:CN
variables Truth Table}, $\ket{true}$ minterms activate the gate
while $\ket{false}$ terms do not.  Under this consideration the
following holds:

\begin{theorem1}\label{theoremquantum}
    A $k-$CN gate is capable of eight characteristic quantum
    operations. $($We consider quantum operations as those that manipulate
quantum phase and non-classical superposition
    states; characteristic has the same meaning as in
    Theorem~\ref{theoremclassical}.$)$
\end{theorem1}
\begin{proof} The proof is constructive:

\emph{Case 1:} When activated, quantum gates exhibit phase kickback
when the state of the target is $\ket{0}+e^{- i\varphi}\ket{1}$. The
activating state can have a phase of $+w_x$ or $-w_x$. Furthermore,
a non-activating state can have a phase of $+w_x$ or $-w_x$ and of
course, nothing should happen when acted on by the $k-$CN gate.

\emph{Case 2:} (The opposite of Case 1.) The alternative case is
that the target acts on state $\ket{0}+e^{+ i\varphi}\ket{1}$.  As
before, the activating and non-activating states can have phases of
$+w_x$ or $-w_x$. Nothing should happen under the case of both an
activating and a non-activating state.  This functionality is probed
in four additional tests.

We draw the readers attention now to the table in Fig.~\ref{tab:CN
variables Truth Table} for the illustration of Case $1$ and Case
$2$. Variables $\phi$ and $\varphi$ are set to create states that
are operated on by the $k-$CN gate, these are the combinations of
actions considered. The Proof is concluded by mentioning that, all
the quantum functions of the $k-$CN gate represent one variant of
these eight cases when used in a phase oracle.~$\blacksquare$
\end{proof}

Thus according to Theorems~\ref{theoremclassical}
and~\ref{theoremquantum} in total we need $4+8=12$ non-entangled
tests to identify the function of any $k-$CN gate. 

\section{The Fault Detection Algorithm}\label{sec:Principle Method}

Tests $T_1$, $T_2$, $T_5$ and $T_6$ verify all classical degrees of
freedom.  Tests $T_3$ and $T_4$ verify the phase kickback features
of the oracle.  As a proof of concept the introduced method holds
the test set size to constant six, increasing the complexity
of added stages for tests $T_3$ and $T_4$. This approach helps
better tie classical ideas with quantum test set generation. This is
due to the fact that classically, circuits realizing linear
functions are easy to test due to their high level of
controllability.

\begin{definition}\label{def:Quantum Built-In Self Test Circuit}
    Quantum Build In Self Test Circuit (\emph{QBIST}): A quantum
    circuit designed to test a second quantum circuit; the quantum circuit under test (\emph{QCUT}).  A \emph{QBIST} circuit
    may be built at the input and/or output terminals of the \emph{QCUT}, and the \emph{QBIST} stage is always assumed to contain no errors.
\end{definition}

Consider the example circuit presented in Fig.~\ref{cir:Quantum
Network Realization of Example Function}.  The analysis given in the
coming subsections begins by generating an input state that turns
all the gates in the network \emph{on} and \emph{off} concurrently.
Denote these tests as
$T_1$ and $T_2$, and their general form on a $k$ variable function
follows:
\begin{description}
  \item[$T_1$:] $(\ket{0}^{\otimes k} + \ket{1}^{\otimes k})\otimes\ket{0}$
  \item[$T_2$:] $(\ket{0}^{\otimes k} - \ket{1}^{\otimes k})\otimes\ket{1}$
\end{description}
The classical equivalent of tests $T_1$ and $T_2$ was given in
Fig.~\ref{fig:Classical test vectors v_1-v_4} (where $T_1$
corresponded to vectors $v_0$ and $v_2$, and $T_2$ corresponded to
both $v_1$ and $v_3$).  Together tests $T_1$ and $T_2$ will be shown
to satisfy Requirements~\ref{Requirement:bit flip},~\ref{Requirement:initialization
fault model},~\ref{Requirement:fadedcontrol},~\ref{Requirement:gates act on full
basis} and~\ref{Requirement:Measurement} in Sec.~\ref{subsec:t1}
and~\ref{subsec:t2}.

Sec.~\ref{subsec:t5 t6} considers tests $T_5$ and $T_6$. These tests
are shown to satisfy Requirement~\ref{Requirement:lost phase} by using the
following states as oracle inputs: $\ket{+}^{\otimes
k}\otimes\ket{+}$ and $\ket{-}^{\otimes k}\otimes\ket{+}$. In both
tests, the state at the controls will not impact the state at the
target, leaving all qubits---ideally---unchanged (since no net
entanglement is generated).

Sec.~\ref{subsec:t3} and~\ref{subsec:t4} investigate the ability of
the network to both attach a relative phase to each activating term
in the superposition and to leave non-activating states unaltered.
This in general is a complex procedure, that in the first case can
be done in two tests denoted as $T_3$ and $T_4$. Test $T_3$ utilizes
state $\ket{+}^{\otimes k}\otimes\ket{-}$ and test $T_4$ utilizes
state $\ket{-}^{\otimes k}\otimes\ket{-}$ as input to the oracle.
However, additional "design-for-test" stages must be added to the
end of the circuit, thereby leading to a deterministic measurement. Tests
$T_3$ and $T_4$ are shown to satisfy Requirement~\ref{Requirement:phase1}. 

\subsection{Test $T_1$: $(\ket{0}^{\otimes k} + \ket{1}^{\otimes k})\otimes\ket{0}$}\label{subsec:t1}
\begin{figure*}[t]
\small{\centerline{  \Qcircuit @C=1.5em @R=.4em  @!R{
 &&&\mbox{\emph{~~~QBIST$_{11}$}}&&&&&& \mbox{\emph{~~~Circuit Under Test}} &&&&&& \mbox{~~~\emph{QBIST$_{12}$}}\\
\lstick{\ket{a}} & \qw      & \qw      & \qw          & \qw       & \targ     & \ctrl{4} & \qw      & \qw      & \qw      & \ctrl{2} & \ctrl{1} & \qw      & \qw      & \qw      & \qw      & \ctrl{1} & \gate{H} & \qw & \qw      & \push{\ket{a}}\gategroup{2}{7}{6}{13}{.7em}{--}\\
\lstick{\ket{a}} & \qw      & \qw      & \qw          & \targ     & \ctrl{-1} & \qw      & \ctrl{3} & \qw      & \qw      & \qw      & \ctrl{2} & \ctrl{1} & \qw      & \qw      & \ctrl{1} & \targ    & \qw      & \qw & \qw      & \push{\ket{a}}\gategroup{2}{2}{6}{6}{1.2em}{^\}}\\
\lstick{\ket{a}} & \qw      & \qw      & \targ        & \ctrl{-1} & \qw       & \qw      & \qw      & \ctrl{2} & \ctrl{1} & \ctrl{2} & \ctrl{2} & \ctrl{2} & \qw      & \ctrl{1} & \targ    & \qw      & \qw      & \qw & \qw      & \push{\ket{a}}\gategroup{2}{14}{6}{18}{.7em}{^\}}\\
\lstick{\ket{a}} & \qw      & \gate{H} & \ctrl{-1}    & \qw       & \dstick{~~~\ket{a}}\qw & \dstick{~~~~~~\ket{\bar{a}}}\qw            & \dstick{~~~~~~\ket{a}}\qw      & \dstick{~~~~~~\ket{\bar{a}}}\qw      & \ctrl{1} & \ctrl{1}   & \dstick{~\ket{a}~~\ket{\bar{a}}~~\ket{a}~~~\ket{\bar{a}}}\qw
& \ctrl{1} & \ctrl{1} & \targ    & \qw      & \qw      & \qw      & \qw & \qw      & \push{\ket{a}}\\
\lstick{\ket{a}} & \qw      & \qw      & \qw          & \qw       &
\qw       & \targ    & \targ    & \targ    & \targ    & \targ    &
\targ    & \targ    & \targ    & \qw      & \qw      & \qw      &
\qw      & \qw & \qw &\push{\ket{a}}}}} \caption{Tests $T_1$ and
$T_2$ (GHZ states): In Test $T_1$, $a=0$ so the circuit starts off
in state: $\ket{0000}$. \emph{QBIST$_{11}$} maps this state to the
oracle's input as: $(\ket{0000}+\ket{1111})\otimes\ket{0}$. In Test
$T_2$, $a=1$ and the input to the oracle is:
$(\ket{0000}-\ket{1111})\otimes\ket{1}$. \emph{QBIST$_{12}$} removes
entanglement and returns the system to a product
state.}\label{cir:t1_on_off_test}
\end{figure*}
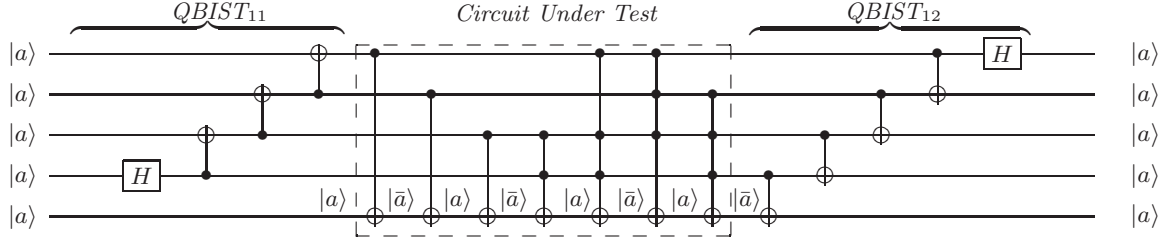

In test $T_1$, all qubits are initialized as:
$\ket{0000}\otimes\ket{0}$. The action of the first
\emph{QBIST$_{11}$} stage (from Fig.~\ref{cir:t1_on_off_test})
creates the following oracle input state:
\begin{equation}
\emph{QBIST$_{11}$}:\ket{0000}\otimes\ket{0}\longrightarrow\left(\ket{0}^{\otimes
k}+\ket{1}^{\otimes k}\right)\otimes\ket{0}.
\end{equation}
The left half of the entangled test sequence is
$\ket{0000}\otimes\ket{0}$. It is clear that for a "\emph{gold
circuit}" not one gate turns on, and the target qubit will be left
untouched. For the right half of the entangled test vector, each
gate in the circuit turns on, and this cycles the $(k+1)^{th}$ qubit
initially starting in $\ket{0}$ back and forth between basis states.
The state of the last qubit after the oracle is
$\ket{0}$.\footnote{If an an even number of gates were present a
slight modification to the final half of the \emph{QBIST}$_{12}$
circuit must be made. This modification is the removal of the first
CN gate at the start of the \emph{QBIST}$_{12}$ acting on the
$(k+1)^{th}$ qubit and controlled by the $k^{th}$ qubit.  In general
for an odd number of gates in a quantum network prior to the final
\emph{QBIST}$_{12}$ stage the circuit will be in state
$\ket{0}^{\otimes k}\ket{0}\pm\ket{1}^{\otimes k}\ket{1}$. The
addition of a CN$_{k,k+1}$ gate removes unwanted entanglement so
that the final qubit will be left in a product state.} The purpose
of \emph{QBIST$_{12}$} is simply to remove the phase induced
entanglement experienced on the top $k$ qubits. The intermittent
states at each stage of the circuit under test $T_1$ are shown in
Fig.~\ref{tab:t1 test pattern impact on stages}.  The final step in
the \emph{QBIST$_{12}$} circuit applies a Hadamard gate to the top
qubit, resulting back in the starting state,
$\ket{0000}\otimes\ket{0}$, thereby completing test $T_1$.  The
complexity of the added CN and H gates needed for test $T_1$ is
$2(k-1)$CN+$2$H.

\begin{figure}[h]\small{\centerline{
\begin{tabular}{|c|c|}
  \hline
  Stage & Action of Stage \\
  \hline
  $in \longrightarrow$ & $\ket{0000}\otimes\ket{0}$ \\
   \hline
  $QBIST_{11}\longrightarrow$ & $(\ket{0000}+\ket{1111})\otimes\ket{0}$ \\
   \hline
  $p_0\longrightarrow$ & $\ket{0000}\ket{0}+\ket{1111}\ket{1}$ \\
   \hline
  $p_1\longrightarrow$ & $\ket{0000}\ket{0}+\ket{1111}\ket{0}$ \\
   \hline
  $p_2\longrightarrow$ & $\ket{0000}\ket{0}+\ket{1111}\ket{1}$ \\
   \hline
  $p_3\longrightarrow$ & $\ket{0000}\ket{0}+\ket{1111}\ket{0}$ \\
   \hline
  $p_4\longrightarrow$ & $\ket{0000}\ket{0}+\ket{1111}\ket{1}$ \\
   \hline
  $p_5\longrightarrow$ & $\ket{0000}\ket{0}+\ket{1111}\ket{0}$ \\
   \hline
  $p_6\longrightarrow$ & $\ket{0000}\ket{0}+\ket{1111}\ket{1}$ \\
   \hline
  $QBIST_{12}\longrightarrow$ & $\ket{0000}\otimes\ket{0}$ \\
  \hline
\end{tabular}}}
\caption{$T_1$ test pattern and impact at each gate in the circuit.
Gates as labeled left to right $p_1$ to $p_6$.}\label{tab:t1 test
pattern impact on stages}
\end{figure}

\subsection{Test $T_2$: $(\ket{0}^{\otimes k} - \ket{1}^{\otimes k})\otimes\ket{1}$}\label{subsec:t2}
No physical change is made to the circuit from
Fig.~\ref{cir:t1_on_off_test}, however the qubits are now
initialized to state $\ket{1111}\otimes\ket{1}$. The outcome is
similar to test $T_1$, the bottom qubit is toggled a total of seven
times resulting in the final state of $\ket{1}$. (\emph{Each gate
that acted on $\ket{0}$ in test $T_1$ now acts on $\ket{1}$ thereby
exhaustively probing every classical input combination of each
$k-$CN gate, seen in Fig.~\ref{cir:t1_on_off_test}.}) The
\emph{QBIST$_{12}$} again disentangles the test responses, resulting
back in the initial state
of $\ket{1111}\otimes\ket{1}$. 

In tests $T_1$ and $T_2$ each node is addressed with both activating
and non-activating states. Furthermore, each qubit is initialized
and measured in both basis states.  Tests $T_1$ and $T_2$ have an
added CN and H gate complexity of $4(k-1)$CN$+4$H. The following
Theorems prove which faults have been detected with tests $T_1$ and
tests $T_2$ and are general for $n$ bit oracles:

\begin{theorem1}\label{theorem:t1 t2 bit flip}
    Either test $T_1$ or test $T_2$ will detect $\sigma_x$ and $\sigma_y$ bit flips at any error location, thus satisfying Requirement~\ref{Requirement:bit flip}.
\end{theorem1}
\begin{proof}
    Tests $T_1$ and $T_2$ both satisfy Requirement~\ref{Requirement:bit flip}. The proof in this section is given for test $T_1$ and is nearly
    identical to the steps taken for test $T_2$. Consider now test $T_1$:

    \emph{Case 1}: The top ($1^{st}$) qubit is flipped:  \emph{QBIST}$_{12}$ receives
    state
    $(\ket{1}\ket{0}^{\otimes(k-1)}\pm\ket{0}\ket{1}^{\otimes(k-1)})$
    as input.  After successive applications of CN$_{i-1,i}$ from $i=k$ to
    $i=2$ the state will be
    $(\ket{11}\ket{0}^{\otimes(k-2)}\pm\ket{01}\ket{1}^{\otimes(k-2)})
    =(\ket{0}\pm\ket{1})\otimes\ket{1}\otimes\ket{0}^{\otimes(k-2)}$.
    Thus, a bit flip impacting the $1^{st}$ bit is detectable on the $2^{nd}$ bit.  
    Given a bit flip impacting any other qubit $q$, ($1<q\leq k$) \emph{QBIST}$_{12}$
    receives
    $(\ket{0}^{\otimes(q-1)}\ket{a}\ket{0}^{\otimes(k-q)}\pm\ket{1}^{\otimes(q-1)}\ket{\bar{a}}\ket{1}^{\otimes(k-q)})$
    as input state.  A similar relation holds such that a bit flip on the $(q-1)^{th}$ bit is detectable on the $q^{th}$
    and possibly the $1^{st}$ bit if the phase is also inverted. For errors impacting any qubit other than the $1^{st}$, both the $q^{th}$ bit as well as
    the $(q+1)^{th}$ (impacted bit) will show the error.

    \emph{Case 2}: Bottom $(k+1)^{th}$ qubit is flipped:  Normally the top $k$ bits and the bottom $(k+1)^{th}$ bits are factorable
    when entering the final \emph{QBIST}$_{12}$ stage.  Assume an even number of gates in the oracle and that instead of
    state: $(\ket{0}^{\otimes k} + \ket{1}^{\otimes
    k})\otimes\ket{0}$ the final \emph{QBIST}$_{12}$ receives the worst case state of
    $\ket{0}^{\otimes k}\ket{0} + \ket{1}^{\otimes
    k}\otimes\ket{1}$.  The final \emph{QBIST}$_{12}$ will not remove the entanglement associated with the $(k+1)^{th}$ bit.  This
    is detectable based on $p$, the probability that a bit flip
    occurred in the computational basis in the first place, satisfying Requirement~\ref{Requirement:bit
    flip}.  This is the only fault that, when deterministically
    present interjects a probabilistic outcome in observability.~$\blacksquare$
\end{proof}

\begin{theorem1}\label{theorem:t1 t2 initialize}
    Together tests $T_1$ and $T_2$ initialize each qubit in both basis states so that Requirement~\ref{Requirement:initialization fault model} is satisfied.
\end{theorem1}
\begin{proof}
    In test $T_1$ the initial state of the register is
    $\ket{0}^{\otimes k}\otimes\ket{0}$ and in test $T_2$ the initial state is
    $\ket{1}^{\otimes k}\otimes\ket{1}$, therefore Requirement~\ref{Requirement:initialization fault model} is satisfied.~$\blacksquare$
\end{proof}

\begin{theorem1}
     Taken together tests $T_1$ and $T_2$ activate all controls concurrently and each
     control is addressed with a non-activating state while the target is separately in basis
     state $\ket{0}$ and next $\ket{1}$ satisfying Requirement~\ref{Requirement:fadedcontrol}.
\end{theorem1}
\begin{proof}
    In tests $T_1$ and $T_2$ the test state prior to application of the oracle
    is $(\ket{0}^{\otimes k} \pm \ket{1}^{\otimes k})\otimes\ket{\bar{a}}$. In both
    tests $T_1$ and $T_2$ the term $\ket{0}^{\otimes k}$ addresses each control with a non-activating
    state, the term $\pm\ket{1}^{\otimes k}$ activates all gates and in both tests the target is in a basis
    state.  This satisfies Requirement~\ref{Requirement:fadedcontrol}.~$\blacksquare$
\end{proof}

\begin{theorem1}\label{theorem:t1 t2 acts on both basis states}
    Taken together tests $T_1$ and $T_2$ force each gate in the circuit to act
    on both basis states, thereby satisfying Requirement~\ref{Requirement:gates act on full basis}.
\end{theorem1}
\begin{proof}
    In both tests $T_1$ and $T_2$ the term $\pm\ket{1}^{\otimes k}$ activates all gates.  Each gate in test $T_1$ that received target input state
    $\ket{a}$ received target input state $\ket{\bar{a}}$ in test $T_2$, thus satisfying Requirement~\ref{Requirement:gates act on full basis}.~$\blacksquare$
\end{proof}

\begin{theorem1}\label{theorem:t1 t2 measured in both basis states}
    After executing test $T_1$ and $T_2$ each qubit will be measured in both basis states, thus satisfying Requirement~\ref{Requirement:Measurement}.
\end{theorem1}
\begin{proof}
    The result of test $T_1$ is $\ket{0}^{\otimes(k+1)}$ and the measured result
    pending the success of test $T_2$ is $\ket{1}^{\otimes(k+1)}$ thus satisfying Requirement~\ref{Requirement:Measurement}.~$\blacksquare$
\end{proof}

\subsection{Tests $T_5$ and $T_6$: $\ket{+}^{\otimes k}\otimes\ket{+}$ and $\ket{-}^{\otimes k}\otimes\ket{+}$}\label{subsec:t5 t6}
The two following tests are simple to conceptualize, as seen in
Fig.~\ref{cir:t5} they have an added gate complexity of $4kH$. When
$a=0$ test $T_5$ generates input state $\ket{++++}\otimes\ket{+}$
and when $a=1$ test $T_6$ generates input state
$\ket{----}\otimes\ket{+}$. Since the eigenvalue of the target state
is $+1$, no change in relative phase should result from propagation
through the quantum circuit and the state of the register should not
become entangled. Theorem~\ref{theorem:t5 t6} proves that test $T_5$
combined with test $T_6$ satisfy Requirement~\ref{Requirement:lost phase} with
an added gate complexity of $4k$H.
\begin{center}
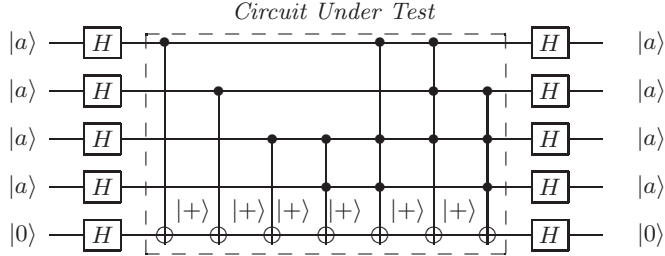
\begin{figure}[h]
\small{\centerline{  \Qcircuit @C=1.4em @R=.7em  {
 &&&&&\mbox{~~\emph{Circuit Under Test}}&&&&  & \\
\lstick{\ket{a}} & \gate{H}      & \ctrl{4} & \qw      & \qw      & \qw      & \ctrl{2} & \ctrl{1} & \qw      & \gate{H} & \qw & \push{\ket{a}}\gategroup{2}{3}{6}{9}{.7em}{--}\\
\lstick{\ket{a}} & \gate{H}      & \qw      & \ctrl{3} & \qw      & \qw      & \qw      & \ctrl{1} & \ctrl{1} & \gate{H} & \qw & \push{\ket{a}}\\
\lstick{\ket{a}} & \gate{H}      & \qw      & \qw      & \ctrl{2} & \ctrl{1} & \ctrl{2} & \ctrl{2} & \ctrl{1} & \gate{H} & \qw & \push{\ket{a}}\\
\lstick{\ket{a}} & \gate{H}      & \dstick{~~~~~~~~~~~\ket{+}~~~~}\qw      & \dstick{~~~~~~~\ket{+}}\qw      & \dstick{~~~~~~~~~~~\ket{+}~\ket{+}}\qw      & \ctrl{1} & \ctrl{1} & \dstick{\ket{+}~\ket{+}}\qw      & \ctrl{1} & \gate{H} & \qw & \push{\ket{a}}\\
\lstick{\ket{0}} & \gate{H}      & \targ    & \targ    & \targ    &
\targ    & \targ    & \targ    & \targ    & \gate{H} & \qw &
\push{\ket{0}}}}}
 \caption{Tests $T_5$ and $T_6$ (Super Tests): Test $\ket{+}^{\otimes k}\otimes\ket{+}$ is first generated ($a=0$, $T_5$) and
 next test $\ket{-}^{\otimes k}\otimes\ket{+}$ is applied ($a=1$, $T_6$).  The target of each $k$-CN gate acts on state $\ket{+}$.
 No entanglement is added in either test, since no relative phase
 change of individual superposition term(s) will occur.}\label{cir:t5}
\end{figure}\end{center}

\begin{theorem1}\label{theorem:t5 t6}
     Together tests $T_5$ and $T_6$ satisfy Requirement~\ref{Requirement:lost phase}.
\end{theorem1}
\begin{proof}
    In both tests $T_5$ and $T_6$ the state of the target qubit is $\ket{+}$. Any
    gate that was activated by a state with an eigenvalue $+1$ in
    test $T_5$ will be activated by a state with an eigenvalue $-1$ in
    test $T_6$.  Relative phase will not change under arbitrary non-activating and activating states
    since the target state has an eigenvalue of $+1$, satisfying Requirement~\ref{Requirement:lost phase}.~$\blacksquare$
\end{proof}

\begin{theorem1}\label{theorem:t5 or t6}
     Either one of tests $T_5$ or $T_6$ detects $\sigma_z$ or $\sigma_y$ phase flips and therefore satisfies Requirement~\ref{Requirement:phase flip}.
\end{theorem1}


\begin{proof}
    Here the Proof is done considering test $T_5$, however the steps are the same as those needed for test
    $T_6$.  Consider state $\ket{+}^{\otimes k}\otimes\ket{+}$,
    this is a product state that may be expanded as:
    $\ket{+}\otimes\cdots\otimes\ket{+}\otimes\ket{+}\otimes\ket{+}\otimes\cdots\otimes\ket{+}$.
    The state of the target is $\ket{+}$ and therefore phase will not make the state
    non-local (with an exception of a phase flip on the $(k+1)^{th}$ bit, in that case the bottom bit will
    deterministically reveal the presence of an error).  Given a $\sigma_z$ fault impacting any qubit, the state
    becomes $\ket{+}\otimes\cdots\otimes\ket{+}\otimes\ket{-}\otimes\ket{+}\otimes\cdots\otimes\ket{+}$.
    In the final stage of  \emph{QBIST}$_{52}$ a Hadamard operation $H^{\otimes(k+1)}$ is applied to the register:
    \begin{eqnarray}\label{eqn:t5}
         &&H^{\otimes(k+1)}\cdot\ket{+}\otimes\cdots\otimes\ket{+}\otimes\ket{-}\otimes\ket{+}\otimes\cdots\otimes\ket{+}\longrightarrow\nonumber\\
         &&~~~~~\ket{0}\otimes\cdots\otimes\ket{0}\otimes\ket{1}\otimes\ket{0}\otimes\cdots\otimes\ket{0}.
    \end{eqnarray}
    Since the $\sigma_z$ bit flip impacts the global state of a qubit, it will be
    seen as a bit flip in the measured state of $T_5$ satisfying Requirement~\ref{Requirement:phase
    flip}.  The proof is concluded mentioning that this result
    coincides with observations drawn in~\cite{Biamonte:05}, (Theorem $2$, $\S~4$).~$\blacksquare$
\end{proof}

\begin{table*}
\small{\centerline{\begin{tabular}{|c|l|c|c|c|c|c|c|c|c|c|}
  \hline
  \emph{Requirements}~($\downarrow$) & \emph{Fault Types Tested}~($\downarrow$) ~~~~~ $-$ ~~~~~ \emph{Tests}~($\rightarrow$) & $T_1$ & $T_2$ & $T_3$ & $T_4$ & $T_5$ & $T_6$ & $T_1\cup T_2$ & $T_3\cup T_4$ & $T_5\cup T_6$\\
   \hline
  Requirement~\ref{Requirement:bit flip} & Any $\sigma_x$ or $\sigma_y$ bit flips occurring? &$\times$&$\times$&  &  &  &  &$\times$&  &  \\
   \hline
  Requirement~\ref{Requirement:phase flip} & Any $\sigma_z$ phase flips occurring? &   &  $\circ$ &  $\circ$ & $\circ$  &$\times$&$\times$&  $\circ$ &  $\circ$ &$\times$\\
   \hline
  Requirement~\ref{Requirement:initialization fault model}&  Is initialization into $\ket{0}$ and $\ket{1}$ O.K.? & $\circ$  &  $\circ$ &   &   &  $\circ$ & $\circ$  &$\times$&$\times$&  \\
   \hline
  Requirement~\ref{Requirement:phase1} & With $\ket{-}$ at target is phase kickback O.K.? &   &   &  $\circ$ & $\circ$  &   &   &   &$\times$&   \\
   \hline
  Requirement~\ref{Requirement:lost phase} & Any phase problems with $\ket{+}$ at the target? &   &   &   &  &  $\circ$ & $\circ$  &   &   &$\times$\\
   \hline
  Requirement~\ref{Requirement:fadedcontrol} & Are the controls activated with $\ket{0}$ and $\ket{1}$? &  $\circ$ & $\circ$  &   &   &   &   &$\times$&   &   \\
   \hline
  Requirement~\ref{Requirement:gates act on full basis} & Gate acts on basis $\ket{0}$ and $\ket{1}$ O.K.? &  $\circ$ &  $\circ$ &   &   &   &   &$\times$&   &   \\
   \hline
  Requirement~\ref{Requirement:Measurement} & Is measurement in $\ket{0}$ and $\ket{1}$ O.K.? & $\circ$  & $\circ$  &   &   &   &   &$\times$&   &   \\
   \hline
\end{tabular}}}\caption{Tests are depicted in columns $3-11$, fault types in column $2$ and Requirements in column $1$.  A given
                        test (column) with table entry $\times$ below it satisfies the Requirement listed in the
                        row corresponding to that $\times$. Entries with $\circ$ inside correspond to tests that cover some, but not all
                        of the faults depicted in the corresponding row.}\label{fig:table of results}
\end{table*}

The classical degrees of freedom for an oracle have been accounted
for in tests $T_1$, $T_2$, $T_5$ and $T_6$ with an added gate
complexity of only $4(k+1)$H+$4(k-1)$CN. The phase kickback features
of the gates in the oracle are verified next in tests $T_3$ and
$T_4$. 

The controllability of a circuit
represents an ability to propagate a specific input vector through a
network, such that it will map a state to a specific fault location.
This represents an added challenge in the case of quantum circuits,
since inputs will become entangled. However, after a
discussion of the upper bounds of tests $T_3$ and $T_4$ in
Sec.~\ref{subsec:Upper Bounds} more controllable test input vectors
are proposed (Sec.~\ref{subsec:Implications}) replacing the added
complexity of these tests with a linear increase in the number of
experiments needed. 

\subsection{Test $T_3$: $\ket{+}^{\otimes k}\otimes\ket{-}$}\label{subsec:t3}
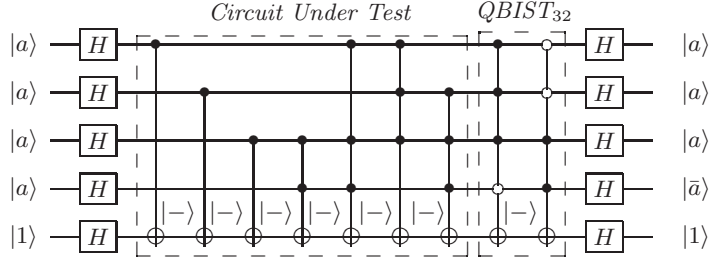
\begin{figure}[h]
\small{\centerline{  \Qcircuit @C=1.2em @R=.7em  {
 &&&&&\mbox{~~\emph{Circuit Under Test}}&&&& \mbox{\emph{~~~~~~QBIST$_{32}$}} & \\
\push{\ket{a}~} & \gate{H}      & \ctrl{4} & \qw      & \qw      & \qw      & \ctrl{2} & \ctrl{1} & \qw      & \ctrl{1}  & \ctrlo{1} & \gate{H} & \qw & \push{\ket{a}}\gategroup{2}{3}{6}{9}{.7em}{--}\\
\push{\ket{a}~} & \gate{H}      & \qw      & \ctrl{3} & \qw      & \qw      & \qw      & \ctrl{1} & \ctrl{1} & \ctrl{1}  & \ctrlo{1}  & \gate{H} & \qw & \push{\ket{a}}\\
\push{\ket{a}~} & \gate{H}      & \qw      & \qw      & \ctrl{2} & \ctrl{1} & \ctrl{2} & \ctrl{2} & \ctrl{1} & \ctrl{1} & \ctrl{1} & \gate{H} & \qw & \push{\ket{a}}\\
\push{\ket{a}~} & \gate{H}      & \dstick{~~~~~\ket{-}}\qw      & \dstick{~~~~~\ket{-}}\qw      & \dstick{~~~~~\ket{-}}\qw      & \ctrl{1} & \ctrl{1} & \dstick{~~~~~\ket{-}~\ket{-}~\ket{-}~~~~~~\ket{-}}\qw      & \ctrl{1} & \ctrlo{1} & \ctrl{1}  & \gate{H} & \qw & \push{\ket{\bar{a}}}\\
\push{\ket{1}~} & \gate{H}      & \targ    & \targ    & \targ    &
\targ    & \targ    & \targ    & \targ    & \targ     & \targ     &
\gate{H} & \qw \gategroup{2}{10}{6}{11}{.7em}{--}& \push{\ket{1}}}}}
 \caption{Circuit Under Test $T_3$ and $T_4$: Test $T_3\longrightarrow\ket{+}^{\otimes k}\otimes\ket{-}$ is first generated ($a=0$, $T_3$) and
 next test $T_4\longrightarrow\ket{-}^{\otimes k}\otimes\ket{-}$ is applied ($a=1$, $T_4$).  Nodes activated with
$\ket{0}$ are denoted as ($\circ$).  \emph{QBIST$_{32}$} removes
entanglement returning the system to a product state and has the
same form in both tests.}\label{cir:t3}
\end{figure}

The goal of test $T_3$ is to verify that phase traverses
correctly amongst all gates. For test $T_3$ the Hadamard gates at
the left of Fig.~\ref{cir:t3} are used to prepare the following
superposition state input on the top $k$ bits:
\begin{eqnarray}\label{eqn:t3}
&&\Longrightarrow\ket{0000}+\ket{0001}+\ket{0010}+\ket{0011}+\ket{0100}+\ket{0101}\nonumber\\
&&+\ket{0110}+\ket{0111}+\ket{1000}+\ket{1001}+\ket{1010}+\ket{1011}\nonumber\\
&&+\ket{1100}+\ket{1101}+\ket{1110}+\ket{1111}
\end{eqnarray}
Observe that Eqn.~\ref{eqn:t3-2} is like a truth table where all the
true minterms of the function have phase factors of $-1$, (see
Fig.~\ref{fig:truth_table}).  This often results in phase induced
entanglement as shown in Eqn.~\ref{eqn:t3-2}.
\begin{eqnarray}\label{eqn:t3-2}
&&\Longrightarrow\ket{0000}+\ket{0001}-\ket{0010}+\ket{0011}-\ket{0100}-\ket{0101}\nonumber\\
&&+\ket{0110} +\ket{0111}-\ket{1000}-\ket{1001}+\ket{1010}+\ket{1011}\nonumber\\
&&+\ket{1100}+\ket{1101}+\ket{1110}-\ket{1111}
\end{eqnarray}
In general, a product (\emph{local}) superposition state may be
written as:
\begin{equation}\label{eqn:exp superposition state}
\pm\bigotimes_{i=0}^{k-1}\left(\ket{0}+a_i\ket{1}\right)
\end{equation}
where any $a_i$ term is either $+1$ or $-1$.  For the state in
Eqn.~\ref{eqn:t3-2} to be expressible as a product state,
Eqn.~\ref{eqn:poly4} must be satisfied:
\begin{equation}\label{eqn:poly4}
    (\ket{0}+a_0\ket{1})(\ket{0}+a_1\ket{1})(\ket{0}+a_2\ket{1})(\ket{0}+a_3\ket{1}).
\end{equation}
Given Eqn.~\ref{eqn:poly4}, any one of $2^i$ ($0 \leq i < k$)
possible choices for $a_i$ results in a local description of the
quantum state (the implications of which will be discussed in
Sec.~\ref{subsec:Upper Bounds}). The general expansion of
Eqn.~\ref{eqn:poly4} leads directly to the generic state:
\begin{eqnarray}\label{eqn:generic}
&&\Longrightarrow\ket{0000}+a_3\ket{0001}+a_2\ket{0010}+a_1\ket{0100}+a_0\ket{1000}\nonumber\\
&&+ a_0\cdot a_1\ket{1100}+a_0 \cdot a_2\ket{1010}+a_0\cdot a_3\ket{1001}\nonumber\\
&&+a_1\cdot a_2\ket{0110}+a_1\cdot a_3\ket{0101}+a_2\cdot a_3\ket{0011} \nonumber\\
&&+a_0\cdot a_1\cdot a_2\ket{1110}+a_0\cdot a_2\cdot a_3 \ket{1011}\nonumber\\
&&+a_0\cdot a_1\cdot a_3\ket{1101}+a_1\cdot a_2\cdot a_3 \ket{0111}\nonumber\\
&&+a_0\cdot a_1\cdot a_2\cdot a_3\ket{1111}
\end{eqnarray}
Comparing Eqns.~\ref{eqn:t3-2} and~\ref{eqn:generic} for the
considered circuit, the system of arithmetic equations given in
Eqn.~\ref{eqn:sys} is obtained. This system is clearly not
specifying a product state since Eqns.~\ref{eqn:t3-2} and
\ref{eqn:generic} matched with Eqn.~\ref{eqn:sys} are inconsistent.
The interfering terms $a_0\cdot a_1\cdot a_2$ and $a_2\cdot a_3$
could be changed for the system to return to a local, product state
description. This may be done by inserting the \emph{QBIST}$_{32}$
circuit given in Fig.~\ref{cir:t3}. \emph{QBIST}$_{32}$ inverts the
phase on terms $\ket{1110}$ and $\ket{0011}$ to $+1$, making the
state factorable as
$(\ket{0}+\ket{1})(\ket{0}+\ket{1})(\ket{0}+\ket{1})(\ket{0}-\ket{1})\otimes\ket{-}$.

\begin{equation}\label{eqn:sys}\left(
  \small{\begin{array}{rr}
    a_0 = -1 & a_1\cdot a_3 = +1 \\
    a_1 = -1 & a_2\cdot a_3 = +1 \\
    a_2 = -1 & a_0\cdot a_1\cdot a_2 = +1 \\
    a_3 = +1 & a_0\cdot a_1\cdot a_3 = -1 \\
    a_0\cdot a_1 = +1 & a_0\cdot a_2\cdot a_3 = +1 \\
    a_0\cdot a_2 = +1 & a_1\cdot a_2\cdot a_3 = +1 \\
    a_0\cdot a_3 = -1 & ~~~a_0\cdot a_1\cdot a_2\cdot a_3 = -1 \\
    a_1\cdot a_2 = +1 & \forall i, a_i\in\{-1,+1\} \\
  \end{array}}
\right)\end{equation}

\subsection{Test $T_4$: $\ket{-}^{\otimes k}\otimes\ket{-}$}\label{subsec:t4}
Test $T_4$ is an exact dual to test $T_3$ and therefore, the needed
\emph{QBIST$_{42}$} stage will have the exact same structure as the
\emph{QBIST$_{32}$} already used. Now the register is initialized
into state $\ket{1111}\otimes\ket{1}$ (by setting $a=1$ in
Fig.~\ref{cir:t3}). The Hadamard operators map this initial state as
follows:
\begin{equation}
\left(H^{\otimes(k+1)}\right)\cdot\left(\ket{1111}\otimes\ket{1}\right)\longrightarrow
\ket{----}\otimes\ket{-}
\end{equation}
and this acts as input to the oracle. The phase of each term is now
opposite when compared with $T_3$.  \emph{QBIST}$_{42}$ inverts the
phase on term $\ket{1110}$ and $\ket{0011}$ to $-1$, making the
state factorable and resulting in this local state description
$(\ket{0}-\ket{1})(\ket{0}-\ket{1})(\ket{0}-\ket{1})(\ket{0}+\ket{1})\otimes\ket{-}$.

Theorem~\ref{theorem:t3 t4} proves that test $T_3$ combined with
test $T_4$ satisfy Requirement~\ref{Requirement:phase1}.  Tests $T_3$ and $T_4$
have a worst case added gate complexity of at most
$\Theta(N-k)+4kH$, where $\Theta$ is a function of the number of
controls needed in the disentanglement stage and the linearity of
the oracle.  

\begin{theorem1}\label{theorem:t3 t4}
     Together tests $T_3$ and $T_4$ satisfy Requirement~\ref{Requirement:phase1}.
\end{theorem1}
\begin{proof}
    In tests $T_3$ and $T_4$ the state of the target is $\ket{-}$. Any
    gate that was activated by a state with eigenvalues $\pm 1$
    during test $T_3$ is activated by a state with eigenvalues $\mp 1$ in
    test $T_4$.  Furthermore, both tests $T_3$ and $T_4$ contain
    non-activating terms, each with opposite eigenvalues.  Tests $T_3$
    and $T_4$ therefore satisfy Requirement~\ref{Requirement:phase1}.~$\blacksquare$
\end{proof}

Table~\ref{fig:table of results} 
provides a concise illustration of the sets of faults entirely
covered by given test(s) (denoted by $\times$) as well as the sets
of faults partially covered by a given test (denoted by $\circ$). We have developed a quantum test algorithm that probes the logical function of each $k-$CN
gate in an oracle.  Now upper bounds on the extraction technique
(\emph{QBIST$_{32}$} circuit stage) will be derived in 
Sec.~\ref{subsec:Upper Bounds}.

\subsection{Upper Bounds for QBIST$_{32}$:}\label{subsec:Upper Bounds}

The concepts of the presented test algorithm are general and
therefore work for any circuit. They do however require the
successful design of the \emph{QBIST$_{32}$}. This design varies
between oracles and has an upper bound of added depth complexity
that depends on the function realized in the oracle. 

\begin{definition}
An affine Boolean function $A_f(x_1,...,x_k)$ , on variables
$x_1,...,x_k$ is any function the takes the form
\begin{equation}
    A_f(x_1,x_2,...,x_k)=c_0\oplus c_1\cdot x_1 \oplus c_2\cdot x_2
    \oplus \cdots\oplus c_k\cdot x_k,
\end{equation}
where $\cdot$ is Boolean AND, $\oplus$ is EXOR (modulo $2$
addition), $c_i\in\{0,1\}$ and $i=0,1,...,n$ are indices of
coefficients. It is easy to see that there exist $2^{k+1}$ affine
functions all of which have checkered cube patterns. A linear
function is any one of the $2^k$ affine functions generated when
coefficient $c_0 = 0$.
\end{definition}

We present the following theorem (\ref{oracle_entanglement})
relating state separability to the function being realized by a
given oracle. 

\begin{theorem1}\label{oracle_entanglement}
Consider oracle $\mathcal{O}$ for which test $T_3$ obtains only
separable (local) measurements (requires no disentanglement).
$\mathcal{O}$ necessarily realizes only affine functions over $k$
variables.
\end{theorem1}

\begin{proof}
The proof is based on the
straightforward generalization of the following example:

Assume input variables $(x_1,x_2,x_3)$.  The expression
\begin{eqnarray}\label{na}
 &&\ket{000}(+1)+\ket{001}a_2+\ket{010}a_1+\ket{011}a_1 \cdot a_2 +\nonumber\\
&& \ket{100}a_0 + \ket{101}a_0\cdot a_2 + \ket{110}a_0\cdot a_1 +\nonumber\\
&&\ket{111}a_0 \cdot a_1 \cdot a_2
\end{eqnarray}
corresponds to a classical truth table with $\prod a_i$ expressions
corresponding to sum-of-product canonical coefficients. Assuming the
encoding
\begin{equation}\label{eqn:encoding}
en(+1)=0,~~~en(-1)=1,
\end{equation}
arithmetic expressions like $a_1\cdot a_2$ are changed to Boolean
values like $en(a_1)\oplus en(a_2)$.  Normally one would consider
the case that $b_0=0$ for linear functions.  Because of global phase
$b_0$ may take either binary value corresponding to all affine
functions on $k$ variables.  It is well known from the canonical SOP
to PPRM conversion method that PPRM = $b_0\cdot 1\oplus(b_0\oplus
b_1)\cdot x_3\oplus(b_0\oplus b_2)\cdot x_2 \oplus (b_0 \oplus b_1
\oplus b_2 \oplus b_3)\cdot x_2\cdot x_3\oplus (b_0\oplus b_2\oplus
b_4\oplus b_6)\cdot x_1\cdot x_2\oplus(b_0\oplus b_4)\cdot x_1
\oplus (b_0\oplus b_1\oplus b_4 \oplus b_5)\cdot x_1\cdot
x_3\oplus(b_0\oplus b_1\oplus b_2\oplus b_3\oplus b_4 \oplus b_5
\oplus b_6\oplus b_7)\cdot x_1\cdot x_2\cdot x_3,$ where $b_i$ are
coefficients of minterms, i.e. $b_0$ is a coefficient of
$\ket{000}$, $b_1$ is a coefficient of $\ket{001}$, etc.  The
minterms of canonical SOP obtain thus the following encoding (symbol
$\cdot$ is arithmetic multiplication)\footnote{This is also called
the polarity table in which one considers a Boolean function over
variables $\{-1,1\}$ instead of $\{0,1\}$.  In this case, XOR
($\oplus$) over $\{0,1\}$ is equivalent to real multiplication over
$\{-1,1\}$.} $b_1 = en(a_2)$, $b_2=en(a_1)$, $b_3=en(a_1\cdot
a_2)=en(a_1)\oplus en(a_2)=b_2\oplus b_1$, $b_4=en(a_0)$,
$b_5=en(a_0\cdot a_2)=en(a_0)\oplus en(a_2)=b_4\oplus b_1$,
$b_7=en(a_0\cdot a_1\cdot a_2)= en(a_0)\oplus en(a_1)\oplus
en(a_2)=b_4\oplus b_2\oplus b_1$.


Applying now the encoding from Eqn.~\ref{eqn:encoding} and
substituting into the above PPRM one obtains PPRM = $b_0\cdot
1\oplus (b_0\oplus b_1)\cdot x_3\oplus (b_0\oplus b_2)\cdot
x_2\oplus [(b_0\oplus b_1\oplus b_2)\oplus (b_2\oplus b_1)]x_2\cdot
x_3\oplus [(b_0\oplus b_2\oplus b_4)\oplus (b_4\oplus b_2)]x_1\cdot
x_2\oplus (b_0\oplus b_4)x_1\oplus [(b_0\oplus b_1\oplus b_4)\oplus
(b_4\oplus b_1)]x_1\cdot x_2\oplus [(b_0\oplus b_1\oplus b_2)\oplus
(b_2\oplus b_1)\oplus (b_4)\oplus (b_4\oplus b_1)\oplus (b_4\oplus
b_2)\oplus (b_4\oplus b_2\oplus b_1)]\cdot x_1\cdot x_2\cdot x_3 =
b_0\cdot 1\oplus (b_0\oplus b_1)x_3\oplus (b_0\oplus b_2)x_2\oplus
(b_0\oplus b_4)\cdot x_1$. Thus, PPRM = $b_0\oplus (b_0\oplus
b_1)x_3\oplus (b_0\oplus b_2)x_2\oplus (b_0\oplus b_4)x_1$ which
corresponds to all affine functions on variables $x_1,x_2,x_3$.
\end{proof}

If oracle $\mathcal{O}$ contains function $f(x_1,...,x_k)$ that is
not affine, a modification to any one of the affine functions
$A_i(x_1,...,x_k)$ must be made. This can be done by adding a
circuit (such as \emph{QBIST}$_{32}(x_1,...,x_k)$) and can be
thought of as EXORing it with some function, like this:
\begin{equation}
f(x_1,...,x_k)\oplus BIST_i(x_1,...,x_k)=A_i(x_1,...,x_k).
\end{equation}
Thus, $f(x_1,...,x_k)=BIST_i(x_1,...,x_k)\oplus A_i(x_1,...,x_k)$.
The general disentanglement procedure is as follows: \\

\begin{enumerate}
  \item Each function $A_i(x_1,...,x_k)\oplus BIST_i(x_1,...,x_k)$ is
realized as an ESOP.
  \item $BIST_i(x_1,...,x_k)$ with the minimum cost is selected.
  \item Function $BIST_i(x_1,...,x_k)$ is added (XORed) after $f$ as
\emph{QBIST}$_{32}$.
\end{enumerate}

\begin{theorem1}
The minimum number of product terms in the ESOP realization of the
BIST circuit ESOP$[$\emph{BIST}$(x_1,...,x_k)\oplus
A_i(x_1,...,x_k)]$ where $A_i$ is an arbitrary affine function on
variables $x_1,...,x_k$ is equal to $p - k$ where $p$ is the minimal
number of product terms in ESOP(\emph{BIST}$(x_1,...,x_k))$.
\end{theorem1}

\begin{proof}
Given is the minimal ESOP, denoted by ESOP(BIST), of function
$BIST(x_1,...,x_k)$. Let $A$ be an arbitrary affine function on
variables $x_1,x_2,...,x_k$ and $c_0\oplus c_1\cdot x_1\oplus ...
c_k\cdot x_k$, where $c_i\in\{0,1\}$.  There are two of these
functions that have the maximum number of variables equaling $k$;
$x_1\oplus x_2\oplus...x_k$ and $1\oplus x_1\oplus
x_2\oplus\cdots\oplus x_k=\bar{x_1}\oplus x_2\oplus ... x_k$.
Assuming that ESOP(BIST) has the minimal number of product terms,
the following cube pair types must not be included in it: $x_i\cdot
x_j\oplus x_i$, $x_i\cdot x_j\oplus x_i\cdot \bar{x_j}$, $x_i\cdot
\bar{x_j}\oplus \bar{x_i}\cdot x_j$, $x_i\cdot x_j\oplus
\bar{x_i}\cdot \bar{x_j}$. The only product terms possible in
ESOP(BIST) are necessarily $x_i$, $\bar{x_i}$, $x_i\cdot x_j$,
$x_i\cdot\bar{x_j}$, $x_i\oplus x_i\cdot x_j\cdot...\cdot x_k$.  If
one writes ESOP(BIST$\oplus A_i$) as ESOP$(BIST)\oplus x_1 \oplus
x_2 \oplus \ldots x_k$ provided all the best merging cases, then all
variables (literals) from $A$ are merged, each of them with some
literal from ESOP(BIST), like this: $x_i\oplus x_i = 0$,
$\bar{x_i}\oplus\bar{x_i}=0$ and $x_i\oplus \bar{x_i}$.  Each of
these cases will decrease the ESOP cost by one.  Merging $x_i$ with
$x_i\cdot x_j = x_i$; $x_i\oplus x_i\cdot x_j = x_i\cdot\bar{x_j}$
will not change the ESOP cost. All other mergings will increase the
cost of the ESOP(\emph{BIST}$\oplus A_i$) with respect to
ESOP(BIST). Thus, the number of terms in the ESOP can be decreased
by no more than $k$.  Observe also that the highest decrease of cost
is when BIST is already an affine function.
\end{proof}



\subsection{Possible Extensions and Applications}\label{subsec:Implications}

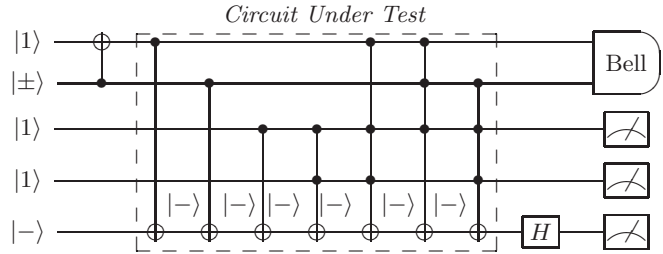
\begin{figure}[h]
\small{\centerline{  \Qcircuit @C=1.4em @R=.7em  {
 &&&&&\mbox{~~\emph{Circuit Under Test}}&&&&  & \\
\push{\ket{1}~} & \targ    & \ctrl{4} & \qw      & \qw      & \qw      & \ctrl{2} & \ctrl{1} & \qw      & \qw      & \multimeasureD{1}{\text{Bell}} \gategroup{2}{3}{6}{9}{.7em}{--}\\
\push{\ket{\pm}~} & \ctrl{-1} & \qw      & \ctrl{3} & \qw      & \qw      & \qw      & \ctrl{1} & \ctrl{1} & \qw      & \ghost{\text{Bell}}\\
\push{\ket{1}~} & \qw      & \qw      & \qw      & \ctrl{2} & \ctrl{1} & \ctrl{2} & \ctrl{2} & \ctrl{1} & \qw      &  \meter\\
\push{\ket{1}~} & \qw      & \dstick{~~~~~~~~~~~\ket{-}~~~~}\qw      & \dstick{~~~~~~~\ket{-}}\qw      & \dstick{~~~~~~~~~~~\ket{-}~\ket{-}}\qw      & \ctrl{1} & \ctrl{1} & \dstick{\ket{-}~\ket{-}}\qw      & \ctrl{1} &  \qw &      \meter\\
\push{\ket{-}~} & \qw      & \targ    & \targ    & \targ    & \targ
& \targ    & \targ    & \targ    &  \gate{H} &      \meter}}}
 \caption{Alternative setup for tests $T_3$ and $T_4$: Test $\ket{0111}\pm\ket{1011}$.  The target of each $k$-CN gate acts on state $\ket{-}$.
 No entanglement is added in either test, since all relative phases
 will result in a product measurement in the Bell basis.}\label{cir:t3a}
\end{figure}

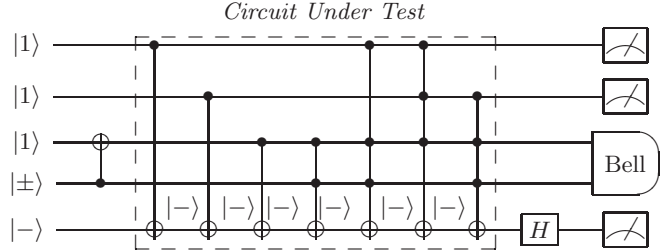
\begin{figure}[h]
\small{\centerline{  \Qcircuit @C=1.4em @R=.7em  {
 &&&&&\mbox{~~\emph{Circuit Under Test}}&&&&  & \\
\push{\ket{1}~} & \qw      & \ctrl{4} & \qw      & \qw      & \qw      & \ctrl{2} & \ctrl{1} & \qw      & \qw      & \meter \gategroup{2}{3}{6}{9}{.7em}{--}\\
\push{\ket{1}~} & \qw      & \qw      & \ctrl{3} & \qw      & \qw      & \qw      & \ctrl{1} & \ctrl{1} & \qw      & \meter\\
\push{\ket{1}~} & \targ    & \qw      & \qw      & \ctrl{2} & \ctrl{1} & \ctrl{2} & \ctrl{2} & \ctrl{1} & \qw      & \multimeasureD{1}{\text{Bell}}\\
\push{\ket{\pm}~} & \ctrl{-1} & \dstick{~~~~~~~~~~~\ket{-}~~~~}\qw      & \dstick{~~~~~~~\ket{-}}\qw      & \dstick{~~~~~~~~~~~\ket{-}~\ket{-}}\qw      & \ctrl{1} & \ctrl{1} & \dstick{\ket{-}~\ket{-}}\qw      & \ctrl{1} & \qw      &  \ghost{\text{Bell}}\\
\push{\ket{-}~} & \qw      & \targ    & \targ    & \targ    & \targ
& \targ    & \targ    & \targ    & \gate{H} &      \meter}}}
 \caption{Alternative setup for tests $T_3$ and $T_4$: Test $\ket{1101}\pm\ket{1110}$.}\label{cir:t3b}
\end{figure}

An alternative approach based on the theory outlined in tests $T_3$
and $T_4$ utilizes highly controllable test vectors.  The growth in
additional circuitry is thus replaced with linear growth in the
number of experiments needed. The total number of
experiments in this second method is $(5 + 4\lceil k/2 \rceil)$.
There is little added growth in circuit complexity. Tests $T_3$ and
$T_4$ are replaced with first repeating the circuitry needed in test
$T_1$. (All replaced tests of course have state $\ket{-}$ at the
target.) Next, starting with the top $2$ qubits
(Fig.~\ref{cir:t3a}), an EPR pair is generated to test the oracle
and mirrored with a measurement in the Bell basis. This is then
moved down all the top $k$ qubits (Fig.~\ref{cir:t3b}) a total of
$2\lceil k/2 \rceil$ times. The EPR generating circuitry is used to
create inputs that are products of state $\ket{01}\pm\ket{10}$ and
$\ket{1}$.  These must be repeated with both positive and negative
versions to satisfy Requirement~\ref{Requirement:phase1}. This results in
something in classical test known as
\emph{walking-a-zero}~\cite{UgurHP} (except quantum mechanics allows
two zeros to be walked at the same time). This alternative approach
however, does not probe the oracle under the types of inputs
experienced when used in a Grover search algorithm. It does however
illustrate that the algorithm can be modified to reduce the
complexity of the stages needed to extract information. Alternative
applications of the methods presented in this paper also exist.

\section{Conclusion}\label{sec:Conclusion}

This work reduced the classical test problem by utilizing
entanglement as a controllability resource. Classically, the lower bound of this circuit class was
found to be $(k+4+2n_e)$ by Reddy~\cite{reddy} (where the $2n_e$
term depends on the function being realized). Quantum effects were used to reduce the test problem 
to a linear growth of $(5 + 4\lceil k/2 \rceil)$ in
experiment count. When testing an oracle, states become non-local
due to the phase change undergone by all true minterms as seen in
tests $T_3$ and $T_4$ in Sec~\ref{subsec:t3} and~\ref{subsec:t4}. It
was shown in Sec.~\ref{subsec:Upper Bounds} that all affine oracles
generate no net entanglement when used as a search oracle, while an
oracle realizing a bent function requires the greatest effort to
disentangle the state and return the system to a local product
state. Since there are $2^{k+1}$ affine functions,
Sec.~\ref{subsec:Upper Bounds} addressed the question of how close
an arbitrary state is to a factorable state with phase terms that
represent the spectrum of an affine function. The distance in many
cases is close, but the upper bound is $\sim \Theta(N - k)$. Linear
and Affine functions are very easy to test when realized quantum
mechanically. Based on the potential limitations highly controllable
test vectors were developed in Sec.~\ref{subsec:Implications} that
do not undergo phase induced entanglement when propagation though a
phase oracle occurs. In a correspondence from Agrawal in $1981$~\cite{agrawal:81}, fault
detection probability was shown to be the highest when the
information output of a circuit is maximized. An information
theoretic approach to quantum fault testing might lead to further
useful insight into the quantum test problem.

%


\end{document}